\documentclass[aps,prl,twocolumn,superscriptaddress,unsortedaddress]{revtex4}

\usepackage{graphicx}
\usepackage{dcolumn}
\usepackage{bm}

\begin{document}

\title{Point contact spectroscopy of Cu$_{0.2}$Bi$_2$Se$_3$ single crystals}

\author{T. Kirzhner}
\affiliation{Physics Department, Technion-Israel Institute of Technology,
Haifa 32000, Israel}

\author{E. Lahoud}
\affiliation{Physics Department, Technion-Israel Institute of Technology,
Haifa 32000, Israel}

\author{K.B. Chaska}
\affiliation{Physics Department, Technion-Israel Institute of Technology,
Haifa 32000, Israel}

\author{Z. Salman}
\affiliation{Paul Scherrer Institute, CH-5232 Villigen PSI, Switzerland}

\author{ A. Kanigel}
\affiliation{Physics Department, Technion-Israel Institute of Technology,
Haifa 32000, Israel}

\date{\today}

\begin{abstract}
We report point contact measurements in high quality single crystals of Cu$_{0.2}$Bi$_2$Se$_3$. We observe three different kinds of spectra:  (1) Andreev-reflection spectra, from which we infer a superconducting gap size of 0.6mV; (2) spectra with a large gap which closes above T$_c$ at about 10K; and (3) tunneling-like spectra with zero-bias conductance peaks. These tunneling spectra show a very large gap of ~2meV ($2\Delta/K_b T_c \sim$14).
\end{abstract}

\maketitle

\section{Introduction}

One of the most promising directions in the fast-growing field of topological phases of matter is the subject of Topological Superconductors ~\cite{tsc1, tsc2}. Closely related to topological insulators, topological superconductors have non-trivial topological properties and in addition are bulk superconductors ~\cite{berg_fu}. The topological superconductors caused a lot of excitement since they are predicted to host Majorana fermions which could be used for quantum computing ~\cite{fu_kane}.

The recently discovered Cu$_x$Bi$_2$Se$_3$ superconductor is a prime candidate for being a topological superconductor ~\cite{hor}. Superconductivity in this compound is achieved by intercalating Cu between the layers of Bi$_2$Se$_3$, which is an archetypical topological insulator. The intercalated Cu acts as an electron dopant leading to superconductivity below $\sim$3.5K.

In this paper, we study the pairing symmetry and the topological features of the Cu$_x$Bi$_2$Se$_3$ crystals using point-contact spectroscopy. Differential conductance measurement of point contacts is a common and effective tool to determine the pairing symmetry of superconductors. Our point-contact measurements showed signs of unconventional superconductivity and a possible realization of topological superconductivity.

\begin{figure}[h!]
\includegraphics[width=8.5cm]{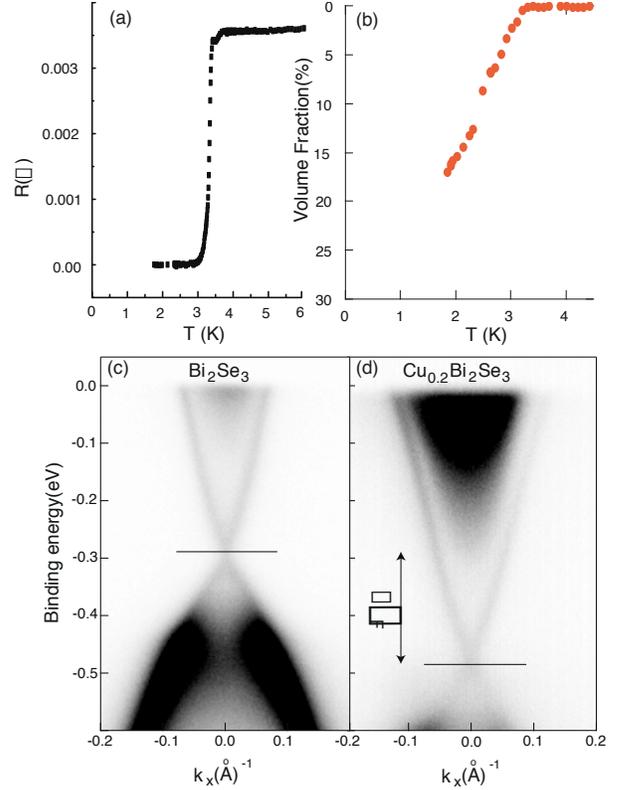}
\caption{(a) Resistivity data showing a zero resistance at $T_c=3.3K$. (b) SQUID magnetization data showing T$_c$=3.3K and a SC volume fraction of 17\% at 1.8K.  (c),(d) ARPES data taken along the $\Gamma-K$ direction at 25K with photon energy of 22eV. The data was measured in the PGM beam-line at the SRC Madison, WI. Panel (c) shows the data for a pristine Bi$_2$Se$_3$ sample and panel (d) for the same Cu$_{0.2}$Bi$_2$Se$_3$ sample used in the point-contact experiments. A large change in the chemical potential can be seen. }
\label{fig1}
\end{figure}

\begin{figure}
\includegraphics[height=6.5cm]{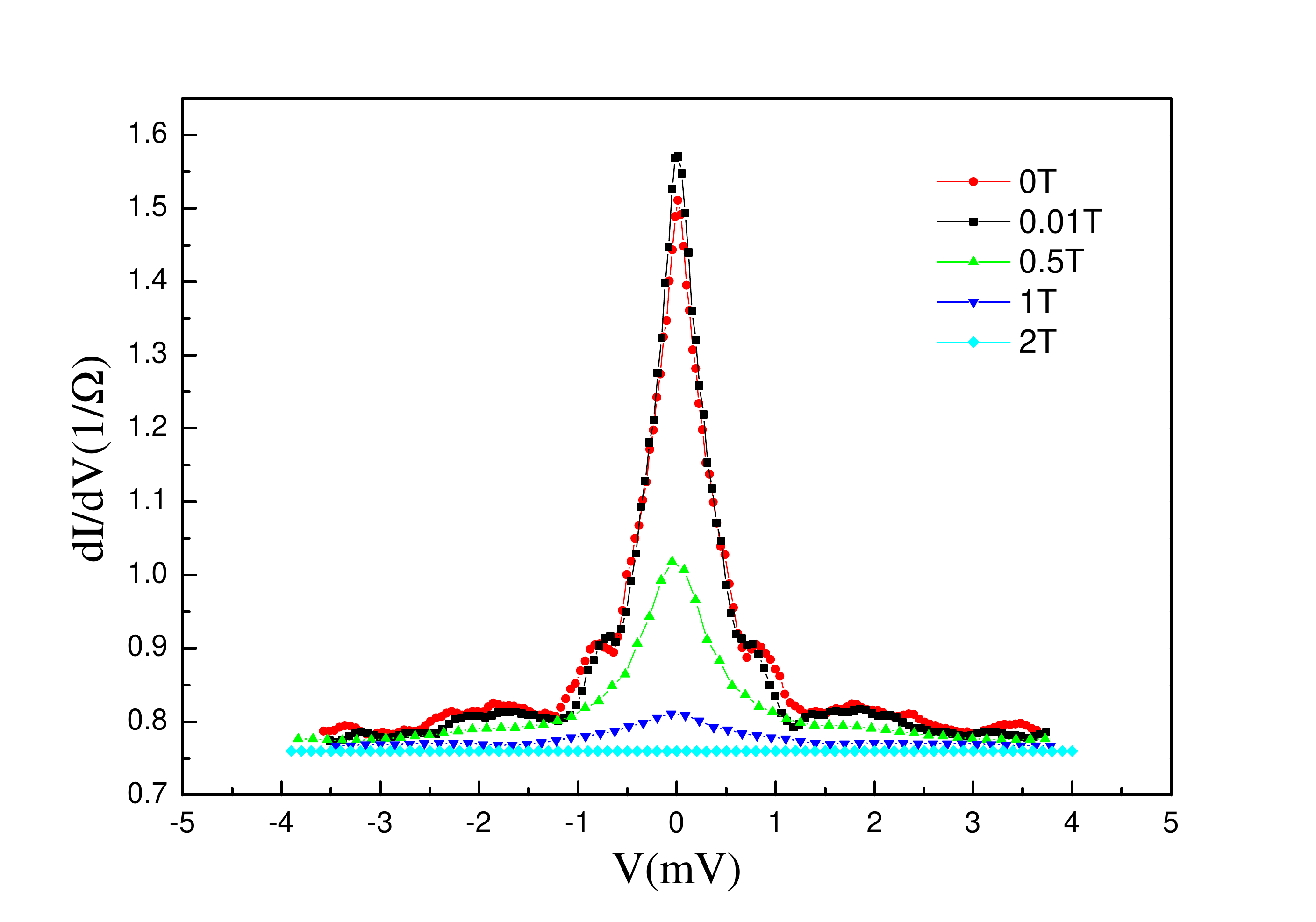}
\caption{\label{fig:arfield2} Differential conductance versus the voltage bias for the Au point contact at temperature of 1.8K at different magnetic fields. }
\end{figure}

\begin{figure*}
\includegraphics[height=6.5cm]{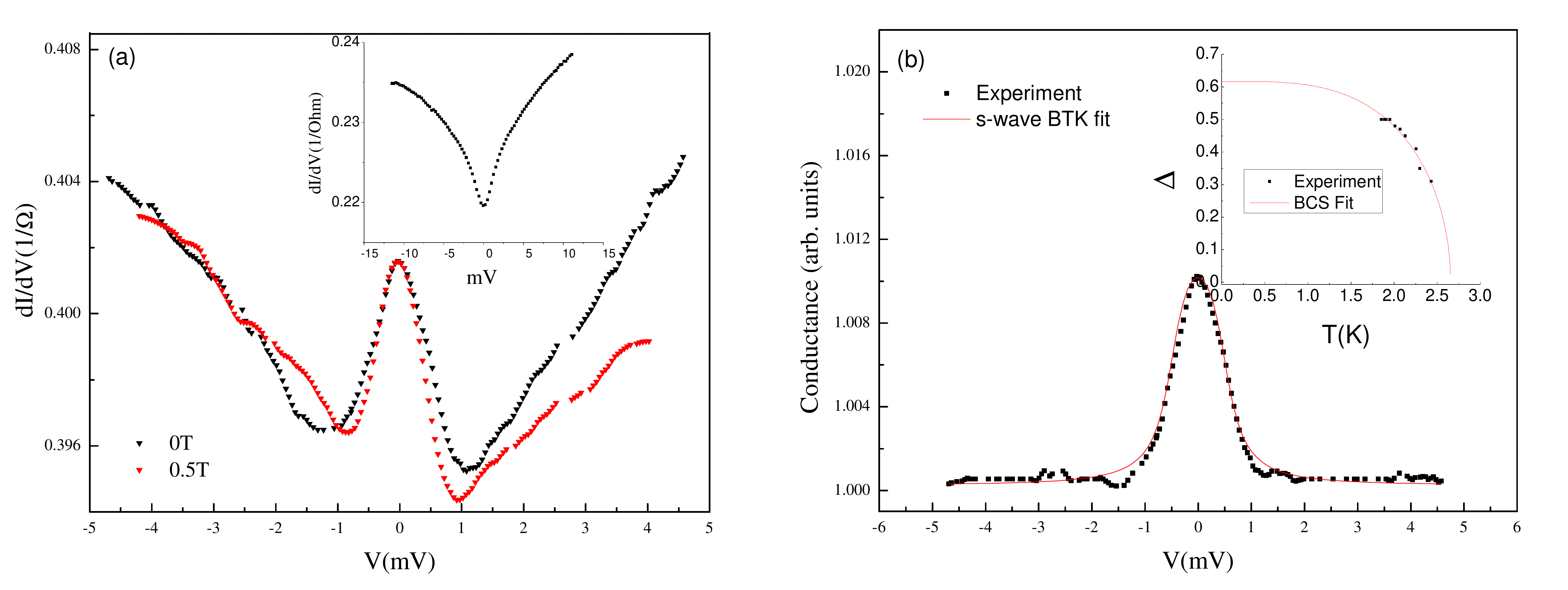}
\caption{\label{fig:arfield} (a) Differential conductance versus the voltage bias for the Au point contact at a temperature of 1.8K at different magnetic fields. Inset: Differential conductance of a point contact on top of a non superconducting spot on the sample. (b) Differential conductance of a point contact with a subtracted background and a fit to an s-wave BTK model with the following parameters: $\Delta=0.5$ mV, and $Z=0$ with a quasiparticle broadening parameter of $\Gamma=0.23$ mV. Inset: The size of the superconducting gap, extracted from the BTK fit as a function of temperature and a fit to the data using a BCS model with a gap size of 0.625mV and $T_c$ of 2.65K . }
\end{figure*}

\section{Experimental details}

High-quality single crystals of Cu$_{0.2}$Bi$_2$Se$_3$ were prepared using the modified Bridgman method. High-purity Cu, Bi and Se, were mixed and sealed in an evacuated fused silica ampoule. The ampoule is warmed up to $850^{\circ}C$ in an inclined tube furnace and then cooled down at a rate of $2.5^{\circ}C$/h to $560^{\circ}C$. The ampoule is held at this temperature for 24 hours and quenched to room temperature. This process is repeated twice with an intermediate grinding. The result is a boule which is easily cleaved normal to the (001) direction of the crystal. Cu$_{0.2}$Bi$_2$Se$_3$ is notorious for being a tricky sample to prepare. This is probably related to the am-bipolar nature of the Cu as a dopant \cite{ambipolar}. We found that the sample is very sensitive to the temperature gradient in the furnace. Resistivity measurement shows a zero resistance at $T_c=3.3K$, as seen in Fig \ref{fig1}(a). Superconducting quantum interference device (SQUID) magnetometry shows these crystals to be superconducting below $T_c=3.3K$. The shielding volume fraction is about 17\% at 2K, as can be seen in Fig ~\ref{fig1}(b).

In order to study the Cu intercalation effect on the electronic properties of our sample we performed angle-resolved photoemission spectroscopy (ARPES) measurements. A comparison between the ARPES spectra of a non-intercalated Bi$_2$Se$_3$ sample grown using the same method and of the superconducting sample is shown in panels (c) and (d) of Fig \ref{fig1}. We show the dispersion along the $K-\Gamma-K$ direction taken with 22eV photons. The n-type doping effect of the Cu intercalation is clearly visible in the figure. The Dirac point sunk down by about 180meV, indicating a large change in the chemical-potential of the sample. We can estimate the carrier concentration in the superconducting (SC) sample to be $2.2\times10^{20}$cm$^{-3}$ . Nevertheless, the surface states are well preserved; we do not see mixing with the conduction band. One can see that the momentum width  of the surface states of the two samples is similar, indicating that the topological protection of the surface state is still intact in the superconducting sample.

We now turn to the main part of this work, the use of point contact spectroscopy to probe the order parameter of the
 Cu$_{0.2}$Bi$_2$Se$_3$ single crystals. A Au tip was used as the point contact. The tip was pressed against the sample surface, and the contact pressure was controlled by mechanical means. The conductance measurements were made using a four-probe technique, and a standard lock-in amplifier technique was used to measure the differential conductance. The quality of our point contacts was verified by measuring Andreev gaps of Nb films using the same system. The surface of the single crystals was imaged using an atomic force microscope after a point-contact measurement, and the contact size was estimated to be a few square microns. The large size of the point contacts increased our chances of contacting superconducting areas. The Cu$_{0.2}$Bi$_2$Se$_3$ single crystals were cleaved using a sticky tape, creating a fresh (001) surface.  Immediately after the cleaving, the sample was cooled down to 1.8K and a differential conductance measurement of the point contacts was made.

\section{Results and discussions}

The resistance of the point contacts was in the range of $1$ to $20\Omega$. About 15\% of the point contacts showed enhancement of the conductance. The zero bias enhancement in all cases appeared below the critical temperature of the bulk superconductivity and the amplitude enhancement ranged between a few percent to about twice of the normal conductance. Application of a 2T field perpendicular to the (001) plane suppressed the zero-bias enhancement in all the point contacts. This result is in line with the reported value for the critical magnetic field in Cu$_{0.2}$Bi$_2$Se$_3$ ~\cite{andoheat}. The small success ratio of the point contacts can be explained by the low volume fraction of the superconductor.

Figure ~\ref{fig:arfield2} shows a differential conductance measurement of a point contact at a temperature of 1.8K. The zero-field curve shows a zero bias peak which starts around 1mV. The amplitude of the peak is about 1.9 times the conductance at higher bias. The Andreev-reflection process is known to enhance the conductance of a superconductor-normal metal junction at energies below the energy gap of the superconductor to a value which is twice the conductance at energies above the gap. Our observation of the zero-bias enhancement seems to fit to a scenario of an Andreev-reflection process. High magnetic field suppresses the zero bias enhancement and at about 2T it disappears.
Zero bias enhancement at point contacts can be also explained by heating effects ~\cite{heating} or by magnetic impurities  ~\cite{magnetic}, which both can be ruled out by application of low magnetic fields. As seen in Fig.~\ref{fig:arfield2} application of 100G did not have much effect on the zero-bias width and height, reinforcing our conclusion that the origin of our zero bias peak is an Andreev-reflection process. Point contacts with different contact resistivity which showed zero bias enhancement, had a common energy scale of 1mV, suggesting that the superconducting energy gap size is around 1mV.

Figure ~\ref{fig:arfield}(a) shows a differential conductance of point contact with a resistance of $2.5\Omega$. In this spectra, the zero bias amplitude is about 1$\%$, which is much lower than in  Fig.~\ref{fig:arfield2}. The low amplitude can be explained by the large size of our point contacts and the low volume fraction of the superconductor. The large point contact consists of many conduction channels, where many of them are just plain metal-metal junctions and only a small part is a metal-superconductor junction, and this may explain the low amplitude of the Andreev peaks. The zero field curve shows an Andreev-reflection like feature, which sits inside an asymmetric V-shaped background. An example spectra of point contact in a non superconducting region with a V-shaped background is shown in the inset of Fig.~\ref{fig:arfield}(a). This background persists well above the critical temperature of the superconductor. As seen in Fig.~\ref{fig:arfield}(a), application of a magnetic field of 0.5T perpendicular to the sample surface, has very little effect on the  height and width of the peak, unlike the magnetic field dependence of the point contact in Fig.~\ref{fig:arfield2}. The variation of the magnetic field dependence in different point contacts is not clear.

\begin{figure*}
\includegraphics[height=6.5cm]{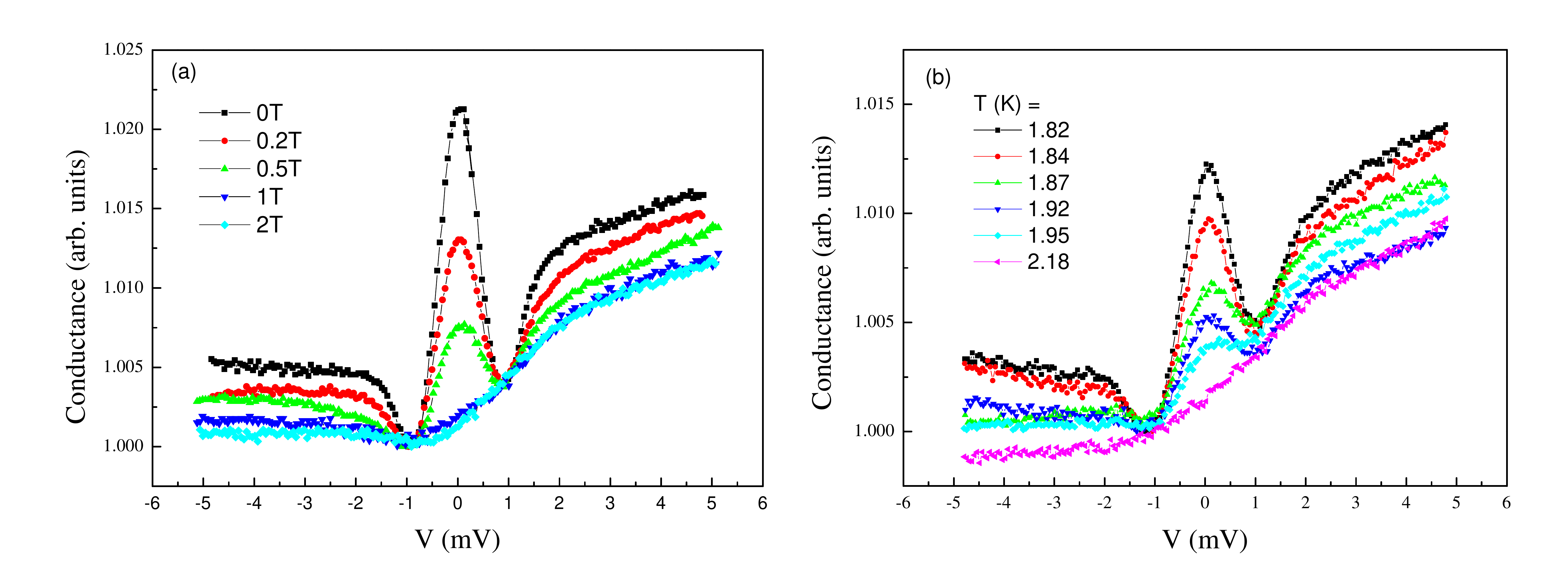}
\caption{\label{fig:zbcp} (a) Differential conductance of a 20 $\Omega$ point contact at different magnetic fields. (b) Temperature dependence of the differential conductance.
}
\end{figure*}

In Fig.~\ref{fig:arfield}(b) we show a differential conductance curve of the same point contact after a background subtraction. We fit the data using an s-wave Blonder-Tinkham-Klapwijk (BTK) ~\cite{BTK} model. The results give an energy gap of about 0.6 mV. The point contacts were measured at a temperature of 1.8K which is high compared to the size of the energy gap, causing broadening of the Andreev peak. We used a quasiparticle broadening parameter of 0.23mV in our fits. This broadening effect makes it impossible to infer from the Andreev data whether a fit to other BTK models with different order parameters, such as d-wave or p-wave, would give better results. Nevertheless, using a simple s-wave model we can get a reasonable estimate of the size of the superconductor energy gap.  In the inset to Fig.~\ref{fig:arfield}(b) we show the SC gap extracted from various curves measured at different temperatures, including a fit to the data using the BCS model. From the fit we find  a gap size of 0.625mV at zero temperature and a critical temperature of 2.6K. The gap value we find is consistent with the results of heat capacity measurements ~\cite{andoheat}.

In Fig.~\ref{fig:zbcp}(a) we show data of a different point contact, this contact had a higher resistance of  $20\Omega$. In this spectra
 we find the characteristic depletion of density of states that appears in junctions with a non transparent barrier (high $Z$ in the BTK model), but without coherence peaks at the gap edges. Within the gap we find a zero bias conductance enhancement.
 One possible interpretation of such conductance structure is that the observed zero-bias conductance peak (ZBCP) is a manifestation of an Andreev bound state originating from the surface of an unconventional superconductor. The peak is not as narrow as in the case of a d-wave superconductor ~\cite{kashiwaya}. The shape of the spectrum with the peak and the dips around it, is similar to the expected spectrum in a chiral p-wave superconductor ~\cite{zbcpchiral}, where the Andreev bound state is dispersive and wide. Fu and Berg showed in Ref~\cite{berg_fu}, that the band structure of Cu$_x$Bi$_2$Se$_3$ may favor an unconventional odd-parity symmetry leading to a time-reversal odd-parity topological superconductor. The proposed superconductor is fully gapped in the bulk, but the surface has gapless surface Andreev bound states. Recently, Hseih and Fu calculated the density of states of the surface states and showed that a dispersive Andreev bound state should appear on the surface of the superconductor, which is consistent with the bound states we see in our point contact data ~\cite{fu}. They also claimed that the zero bound state seen in Cu$_x$Bi$_2$Se$_3$ crystals host Majorana fermions.

 Fig.~\ref{fig:zbcp}(b) shows the temperature dependance of the differential conductance of the same point contact. The position of the dips did not change as the temperature was increased and the peak disappeared at about 2.1K.  The temperature dependence of the gap-size we find here is not BCS-like and can be an indication for an unconventional superconductivity. The height of the ZBCP was suppressed by a magnetic field as seen in Fig.~\ref{fig:zbcp}(a), and it was suppressed completely at 2T. A strong magnetic field did not change the position of the dips, ruling out heating effects ~\cite{heating}.

\begin{figure}
\includegraphics[height=6.5cm]{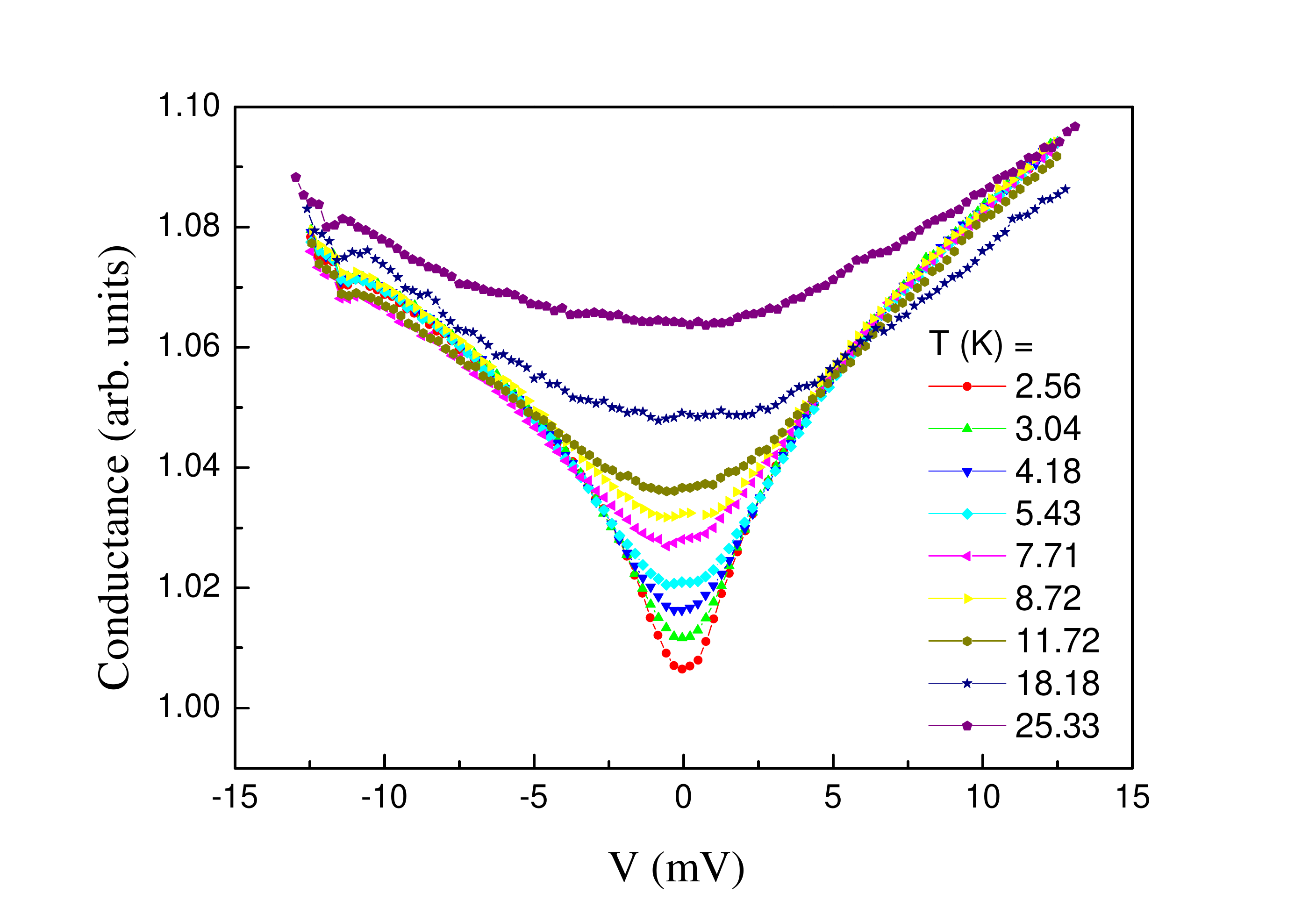}
\caption{\label{fig:pg} Differential conductance measurements made at high temperatures, showing a large gap.
}
\end{figure}

 The dips in the measured spectra start at ~2mV, suggesting a very large gap compared to the results inferred from the Andreev-reflection spectra. This gap size is very large compared to the $T_c$ of the crystal, giving a gap to critical temperature ratio of $2\Delta/K_b T_c \sim$14 (taking $T_c=3.3K$). Such a large gap size makes the scenario of a ZBCP coming from unconventional superconductivity less likely.
 The observed ZBCP could, in principle, be the result of two parallel channels, one with a transparent barrier with an Andreev peak and one channel with a larger gap. The width of the ZBCP in Fig.~\ref{fig:zbcp} is very close to the width of the Andreev peaks we have shown in previous figures, strengthening the proposed scenario. We had observed many spectra with large gaps in many point contacts, with size ranging from 2mv to 5mv. In Fig.~\ref{fig:pg} we see an example of such large gap with a size of ~5mV, which closes above 10K. In Fig.~\ref{fig:zbcp} we can observe that at high magnetic fields and at high temperature the ZBCP vanishes, revealing a large gap of 2mV. The sharpness of the revealed gap is less pronounced than in the low temperature regime, not ruling out completely the existence of a ZBCP with dips. Such large gaps can also be seen in the background of Fig.~\ref{fig:arfield}(a).

Our results partly agree with the recent results of Sasaki {\it et al.} ~\cite{andozbcp}, but there are few differences.  First, they measured point contacts of Cu$_x$Bi$_2$Se$_3$ crystals which were made using an electrochemical intercalation method. Second, they observed a ZBCP which vanished at a lower temperature of ~1K and the gap size was smaller ($\Delta\simeq0.6mV$) compared to the gap size we have observed in our ZBCP spectra. The inferred gap in their measurements agrees with our Andreev spectra.

\section{Conclusions}

To summarize, we measured the differential conductance of single crystals of Cu$_{0.2}$Bi$_2$Se$_3$ using a point contact. We find three different kinds of spectra. The first kind consists of an Andreev-like peak, a fit to a BTK model givne a gap size of 0.6meV, which is in agreement with the observations of other groups. The second kind are spectra showing a large gap, 2-5meV, which persist above $T_c$. The third kind, which is the most interesting, are tunneling-like spectra with a ZBCP. The position of the dips in the tunneling spectra implies a very large gap size. These spectra can be the result of an Andreev bound state in an unconventional superconductor, but can also be a result of two parallel conductance channels: an Andreev-like one and one having a large gap. Using a scanning tunneling microscope, it will probably be possible to distinguish between these two scenarios.

\begin{acknowledgments}
We are grateful to G. Koren for helpful discussions and for the critical reading of this manuscript.
This work was supported by the Israeli Science Foundation.
 The Synchrotron Radiation Center is supported by NSF Grant No. DMR-0084402.
\end{acknowledgments}

\bibliography{majorana}

\end{document}